\documentclass[sigconf]{acmart}

\usepackage{algorithm, algorithmic, amsmath, bm, subcaption, cleveref, multirow, makecell, mathtools, graphicx, balance}
\usepackage[normalem]{ulem}
\useunder{\uline}{\ul}{}

\AtBeginDocument{%
  }

\setcopyright{acmlicensed}
\copyrightyear{2018}
\acmYear{2018}
\acmDOI{XXXXXXX.XXXXXXX}

\acmConference[Conference acronym 'XX]{Make sure to enter the correct
  conference title from your rights confirmation emai}{June 03--05,
  2018}{Woodstock, NY}
\acmISBN{978-1-4503-XXXX-X/18/06}

\newcommand{\beqa}{\begin{eqnarray}}
\newcommand{\eeqa}{\end{eqnarray}}
\newcommand{\beq}{\begin{equation}}
\newcommand{\eeq}{\end{equation}}
\newcommand{\ben}{\begin{enumerate}}
\newcommand{\een}{\end{enumerate}}
\newcommand{\bit}{\begin{itemize}}
\newcommand{\eit}{\end{itemize}}
\newcommand{\bi}{\begin{itemize} \item}
\newcommand{\ei}{\end{itemize}}

\newcommand{\begindef}{\begin{Definition} \rm}
\newcommand{\beginexa}{\begin{Example} \rm}
\newcommand{\beginthe}{\begin{Theorem} \rm}
\newcommand{\beginpro}{\begin{Proposition} \rm}
\newcommand{\beginlem}{\begin{Lemma} \rm}
\newcommand{\begincon}{\begin{Conjecture} \rm}
\newcommand{\begincor}{\begin{Corollary} \rm}

\newcommand{\eat}[1]{}

\def\papernumber #1 raised #2 {
\vspace{-#2}
\vbox to 0pt{\hfill\framebox{\bf Paper Number #1}}
\vspace{#2}
}

\def\T{{\scriptscriptstyle\mathsf{T}}}
\def\x{\mathbf{x}}
\def\X{\mathbf{X}}
\def\s{\mathbf{s}}
\def\S{\mathbf{S}}

\def\algname{\textsc{InterFormer}}

\begin{document}

\title{\algname: Effective Heterogeneous Interaction Learning for Click-Through Rate Prediction}


\author{Zhichen Zeng$^{*1}$\, Xiaolong Liu$^{*3}$\, Mengyue Hang$^{*2}$\, Xiaoyi Liu$^{*2}$\, Qinghai Zhou$^2$\, Chaofei Yang$^2$\, Yiqun Liu$^2$\, Yichen Ruan$^2$\,Laming Chen$^2$\, Yuxin Chen$^2$\, Yujia Hao$^2$\, Jiaqi Xu$^2$\, Jade Nie$^2$\, Xi Liu$^2$\, Buyun Zhang$^2$\, Wei Wen$^2$\, Siyang Yuan$^2$\, Hang Yin$^2$\, Xin Zhang$^2$\, Kai Wang$^2$\, Wen-Yen Chen$^2$\, Yiping Han$^2$\, Huayu Li$^2$\, Chunzhi Yang$^2$\, Bo Long$^2$\, Philip S. Yu$^3$\, Hanghang Tong$^1$\, Jiyan Yang$^2$\\
$^1$ University of Illinois Urbana-Champaign, $^2$ Meta AI, $^3$ University of Illinois Chicago \\
\small\texttt{\{zhichenz, htong\}@illinois.edu, \{xliu262, psyu\}@uic.edu,
\{hangm, xiaoyliu, chocjy\}\allowbreak@meta.com}}
\renewcommand{\shortauthors}{Zhichen Zeng et al.}

\begin{abstract}
  A clear and well-documented \LaTeX\ document is presented as an
  article formatted for publication by ACM in a conference proceedings
  or journal publication. Based on the ``acmart'' document class, this
  article presents and explains many of the common variations, as well
  as many of the formatting elements an author may use in the
  preparation of the documentation of their work.
\end{abstract}


\received{20 February 2007}
\received[revised]{12 March 2009}
\received[accepted]{5 June 2009}

\begin{abstract}
Click-through rate (CTR) prediction, which predicts the probability of a user clicking an ad, is a fundamental task in recommender systems.
    The emergence of heterogeneous information, such as user profile and behavior sequences, depicts user interests from different aspects.
    A mutually beneficial integration of heterogeneous information is the cornerstone towards the success of CTR prediction.
    However, most of the existing methods suffer from two fundamental limitations, including (1) \textit{insufficient inter-mode interaction} due to the unidirectional information flow between modes, and (2) \textit{aggressive information aggregation} caused by early summarization, resulting in excessive information loss.
    To address these limitations, we propose a novel module named \algname\ to learn heterogeneous information interaction in an \textit{interleaving} style.
    To achieve better interaction learning, \algname\ enables bidirectional information flow for mutually beneficial learning across different modes.
    To avoid aggressive information aggregation, we retain complete information in each data mode and use a separate Cross Arch for effective information selection and summarization.
    \algname\ has been deployed across multiple platforms at Meta Ads, achieving 0.15\% performance gain and 24\% QPS gain compared to prior state-of-the-art models and yielding sizable online impact.
\end{abstract}
\vspace{-5pt}

\maketitle

\section{Introduction}\label{sec:intro}

Click-through rate (CTR) prediction, which predicts the probability of a user clicking an item, is the fundamental task for various applications such as online advertising and recommender systems~\cite{zhang2022dhen, zhang2024wukong}.
The quality of CTR prediction significantly influences company revenue and user experience, drawing extensive attention from academia and industry~\cite{zhou2018deep,zhou2019deep,liu2024collaborative,liang2025external,yoo2024ensuring,liu2023class,liu2024aim,liu2025breaking}.
For example, in ad bidding, CTR prediction helps advertisers optimize the bids and target the most receptive audiences. In content recommendation, it enables platforms to suggest relevant content to users.

To achieve better CTR prediction, it is crucial to capture the user interests in the evolving environment~\cite{zhou2018deep,lyu2020deep,wang2019sequential}.
The abundance of heterogeneous information presents both opportunities and challenges.
On the one hand, heterogeneous information depicts user interests from different aspects, providing diverse context~\cite{zhang2017joint}.
For instance, the non-sequence features, e.g., user profile and context features, offer a static view on general user interests, while behavior sequences provide substantial information for modeling dynamic user interests~\cite{wang2019sequential}.
On the other hand, the heterogeneous nature of the data requires different modeling approaches and careful integration across different information modes~\cite{zhang2017joint}.
For example, while modeling interactions among non-sequential information is critical to personalized recommendation~\cite{rendle2010factorization,lian2018xdeepfm,wang2021dcn}, capturing sequential dependencies is the major focus for user behavior modeling~\cite{sun2019bert4rec, chen2019behavior}.

Most of the existing CTR prediction models fall into two categories, including \textit{non-sequential} and \textit{sequential models}.
Non-sequential models focus on learning informative embeddings through feature interaction via inner-product~\cite{lian2018xdeepfm,sun2021fm2}, MLP~\cite{wang2017deep,wang2021dcn} and deep structured semantic model~\cite{huang2013learning,elkahky2015multi}, but ignore the sequential information in user behaviors.
Sequential models, in contrast, employ additional modules like CNN~\cite{tang2018personalized}, RNN~\cite{sun2019bert4rec,zhou2018deep} and Attention modules~\cite{lyu2020deep,zhou2019deep,zhai2024actions}, to capture the sequential dependencies in user behaviors.
Promising as it might be, existing sequential methods mostly employ a unidirectional information flow, where the non-sequential information is used to guide sequence learning, while the reverse information flow from sequence to non-sequence is largely ignored, resulting in insufficient inter-mode interaction. For example, the non-sequential information captures long-term interests, while the sequential information reveals momentary interests, such as a sudden focus on a specific category of products, which can enhance the non-sequential context with immediate preference.
Besides, due to the computational challenges of performing interaction learning among numerous non-sequence features and lengthy sequences, aggressive feature aggregation, e.g., sequence summation~\cite{zhou2018deep}, pooling~\cite{xiao2020deep}, and concatenation~\cite{zhou2019deep}, is often performed in the early stages, inevitably leading to excessive information loss.

In light of the above limitations, we propose a novel heterogeneous interaction learning module named \algname, whose ideas are two-fold.
To avoid insufficient inter-mode interaction, we enable bidirectional information flows between different modes, such that non-sequence and sequence learning are performed in an interleaving style.
Specifically, to learn context-aware sequence embeddings, non-sequence summarization guides sequence learning via Personalized FeedForward Network (PFFN) and Multihead Attention (MHA)~\cite{vaswani2017attention}.
To learn behavior-aware non-sequence embeddings, sequence summarization instructs non-sequence learning via an interaction module.
To mitigate aggressive information aggregation, we adopt MHA for effective information selection such that the one-to-one mappings between input and output tokens are retained.
Note that our framework is compatible with various interaction learning models like DCNv2~\cite{wang2021dcn}, DHEN~\cite{zhang2022dhen}, etc.

The main contributions of this paper are summarized as follows:
\vspace{-2pt}
\begin{itemize}
    \item \textbf{Challenges.} We identify two key bottlenecks of heterogeneous interaction learning, namely insufficient inter-mode interaction and aggressive information aggregation.
    \item \textbf{Model Design.}
    A novel heterogeneous interaction learning framework named \algname\ is proposed for effective feature interaction and selective information aggregation. To our best knowledge, we are the first to address the mutual benefits in heterogeneous interaction learning.
    \item \textbf{Experiments and Analysis.} Extensive experiments suggest that \algname\ achieves up to 0.14\% AUC improvement on benchmark datasets and 0.15\% Normalized Entropy (NE) gain on industrial dataset. In addition, the internal deployment at Meta suggests that \algname\ exhibits promising scaling results, achieving 0.15\% performance gain and 24\% QPS gain compared to SOTA CTR models, and yielding sizable online impact.
\end{itemize}

The rest of the paper is organized as follows. Section~\ref{sec:related} briefly reviews the recent works on interaction learning. Section~\ref{sec:prelim} summarizes the preliminaries, and section~\ref{sec:method} introduces our proposed \algname. Extensive experiments and analyses are carried out in Section~\ref{sec:exp}. We conclude our paper in Section~\ref{sec:con}.

\section{Related Works}\label{sec:related}
In the era of big data and AI \cite{yan2021dynamic,yan2021bright,yan2023trainable,yan2023reconciling,yan2022dissecting,yan2024pacer,yan2024topological,yan2024thegcn,yanred,xu2024slog,yu2025joint,yu2025planetalign,zeng2024graph,zeng2025pave,zeng2023parrot,zeng2023generative,zeng2024hierarchical,bao2024matcha,xu2024discrete,lin2024duquant,lin2025quantization,lin2025toklip}, 
recommender systems have drawn significant attention, exerting profound influence on company revenue and user experience~\cite{wang2018acekg,wang2023networked,wang2017deep,wang2021dcn,li2022unsupervised,li2024large,lee2019set,jing2024sterling,jing2022coin,yang2024simce,wang2023noisy}.
In this section, we review the related works on CTR prediction, including non-sequential and sequential methods.

\vspace{-3pt}
\subsection{Non-Sequential Methods}
The vast majority of non-sequential models is built upon the idea of Factorization Machine (FM)~\cite{rendle2010factorization, lian2018xdeepfm, sun2021fm2}, which models the user-item interaction by linearly combining their low-dimensional embeddings~\cite{zhang2019deep}.
\cite{rendle2010factorization} is the very first FM model to capture pairwise interactions.
To model high-order interactions, different methods combine FMs with deep neural networks, where FMs learn low-order interaction via pairwise operation and neural networks learn high-order interactions via deep architectures. For example, MLP~\cite{lian2018xdeepfm,yang2017bridging, wang2021dcn, sun2021fm2, zhou2020can} captures high-order interactions via dense connections between features, and Attention mechanism~\cite{song2019autoint, xiao2017attentional, xin2019cfm} learns more complex embeddings through linear combination.
These deep learning-based approaches enable end-to-end training without hand-craft feature designs, and are capable of handling heterogeneous information like text, image and video~\cite{zhang2019deep}.
Besides, recent works address the scaling law in recommendation, where DHEN~\cite{zhang2022dhen} ensembles multiple interaction modules and Wukong~\cite{zhang2024wukong} stacks FMs to learn a hierarchy of interactions.
Promising as they are, non-sequential models fail to capture the sequential dependencies in user behaviors, resulting in suboptimal solutions.

\vspace{-3pt}
\subsection{Sequential Methods}
With the recent emergence of sequential information, e.g., user interaction history, in recommender systems, extensive efforts have been made to capture the evolving user interest.
A key challenge behind sequential methods is to combine the sequential information with non-sequential information in a mutually beneficial manner.
Markov models~\cite{shani2005mdp,he2016fusing,yang2020hybrid} consider sequential data as a stochastic process over discrete random variables, but the oversimplified Markovian property limits the model capability in capturing long-term sequential dependencies~\cite{quadrana2018sequence}.
To model the long-term dependencies, RNN and Attention mechanism are employed as the backbone module for many sequential methods. For example, various attention-based networks are designed~\cite{zhou2019deep, lyu2020deep}, and Transformer~\cite{devlin2018bert, vaswani2017attention,lin2025cats} is adopted for sequential modeling~\cite{sun2019bert4rec,chen2019behavior}.
Besides, to model multi-faceted user interests, \cite{xiao2020deep, han2024efficient} propose to capture multiple user interests from multi-behavior sequences.
More recently, TransAct~\cite{xia2023transact} adopt a hybrid ranking model to combine real-time user actions for immediate preference and batch user representations for long-term interests.
LiRank~\cite{borisyuk2024lirank} improves CTR prediction at LinkedIn by ensembling multiple interaction modules, accelerated by quantization and vocabulary compression.
CARL~\cite{chen2024cache} achieves fast inference on large-scale recommendation in Kuaishou by utilizing cached results, monitored by a reinforcement learning-based framework.
While most of the existing sequential methods leverage non-sequential information for personalized sequence modeling, sequential information is rarely explored for non-sequence learning.
We believe such unidirectional design limits the expressiveness of the learned embeddings, and a bidirectional information flow between different modes is the key towards better heterogeneous interaction learning.

\section{Preliminaries}\label{sec:prelim}
Table~\ref{tab:sym} summarizes the main symbols and notations used throughout the paper.
We use bold uppercase letters for matrices (e.g., $\mathbf{X}$), bold lowercase letters for vectors (e.g., $\mathbf{x}$), and lowercase letters for scalars (e.g., $n$).
The element at the $i$-th row and $j$-th column of a matrix $\mathbf{X}$ is denoted as $\mathbf{X}(i,j)$. The transpose of $\mathbf{X}$ is denoted by the superscript $\T$ (e.g., $\mathbf{X}^\T$).
We use superscript $u$ to denote users, and subscripts $i$ and $t$ to denote item and timestamp, respectively (e.g., $y^u_{i_t}$).
We use $\mathbf{x}_j^{(l)}$ to denote the $j$-th non-sequence feature of the $l$-th layer, and $\mathbf{s}_t^{(l)}$ to denote the sequence feature of the $l$-th layer at timestamp $t$. We consider the scenario with $m$ dense features, $n$ sparse features, and $k$ sequence features of length $T$. We use $d$ to denote the embedding dimension of the CTR model.
\begin{table}
  \caption{Symbols and notations.}
  \vspace{-10pt}
  \label{tab:sym}
  \begin{tabular}{@{}cc@{}}
    \toprule
    Symbol &Definition\\
    \midrule
    $\mathcal{U},\mathcal{I}$ & user set and item set\\
    $\x_j^{(l)}$ &$j$-th non-sequence feature of layer $l$\\
    $\s_t^{(l)}$ &sequence feature of layer $l$ at time step $t$\\
    $m, n, k$ &number of dense, sparse and sequence features\\
    $T,d$ &sequence length and embedding dimension\\
    $\odot$ &Hadamard product\\
    $\langle\cdot,\cdot\rangle$ &inner product\\
    $[\cdot\|\cdot]$ & horizontal concatenation of vectors\\
    \bottomrule
\end{tabular}
\end{table}

\vspace{-5pt}
\subsection{Click-Through Rate (CTR) Prediction}
CTR prediction estimates the probability of a user clicking on an item given heterogeneous information such as static context and behavior sequences. Formally, given a user set $\mathcal{U}$ and an item set $\mathcal{I}$, the interaction sequence of a user $u\in\mathcal{U}$ is defined as $S^u = [i^u_1,i^u_2,\dots, i^u_T]$, where $i^u_t \in \mathcal{I}$ is the item interacted at time step $t$ by user $u$.
We aim to estimate the probability of a user clicking on a new item $i^u_{T+1}$, denoted as $y^u_{i_{T+1}}$, given the historical interaction sequence $S^u$. Formally, we seek to learn a function $f: \mathcal{U} \times \mathcal{I}\times \mathcal{S} \rightarrow [0, 1]$, where $\mathcal{S}$ is the set of all possible sequences, such that:
\begin{equation}\label{eq:ctr}
    P(y^u_{i_{T+1}} = 1 | u, i_{T+1}, S^u; \theta) = f(u, i_{T+1}, S^u; \theta).
\end{equation}
To optimize the model, the cross-entropy loss, which measures the difference between predicted and groundtruth click probabilities, is commonly employed as the objective function. Since the formulation in Eq.~\eqref{eq:ctr} incorporates temporal dynamics in user behaviors, it potentially leads to better CTR predictions than non-sequential methods that ignore the sequential nature of user interactions.

\vspace{-2pt}
\subsection{Feature Interaction}
Interaction learning, which aims to capture the complex relationships between different features, is the key towards the success of CTR prediction. We briefly introduce three prominent feature interaction modules: inner product, DCNv2~\cite{wang2021dcn} and DHEN~\cite{zhang2022dhen}.

\noindent\textbf{Inner Product-based Interaction.}
Given input feature vector $\x\in\mathbb{R}^d$, the inner product-based interaction, exemplified by Factorization Machines (FM) \cite{rendle2010factorization}, learns a latent vector $\mathbf{v}_j \in \mathbb{R}^r$ for each feature $j$, whose inner product $\langle\mathbf{v}_j,\mathbf{v}_k\rangle$ describes the interaction strength between features $\x(j)$ and $\x(k)$. Hence, the second-order feature interactions can be modeled as:
\begin{equation*}
    f_{FM}(\mathbf{x}) = \sum_{j=1}^d \sum_{k=j+1}^d \langle \mathbf{v}_j, \mathbf{v}_k \rangle \x(j)\x(k) + \sum_{j=1}^d w_j\x(j) + w_0.
\end{equation*}

\noindent\textbf{Deep \& Cross Network (DCNv2) ~\cite{wang2021dcn}.}
The DCNv2 model combines a cross network for explicit feature interactions with a deep neural network for implicit feature interactions. Given input feature vector $\mathbf{x}^{(0)} \in \mathbb{R}^d$, the $l$-th layer of the cross network models the second order interaction as follows:
\begin{equation*}
    \mathbf{x}^{(l+1)} = \mathbf{x}^{(0)}\odot\left(\mathbf{w}^{(l)}\mathbf{x}^{(l)}  + \mathbf{b}^{(l)}\right) + \mathbf{x}^{(l)}.
\end{equation*}
DCNv2 further combines the stacked cross layers for explicit interactions with a deep layer for implicit interactions, enabling an output enriched with comprehensive information.

\noindent\textbf{Deep Hierarchical Ensemble Network (DHEN)~\cite{zhang2022dhen}.}
DHEN learns a hierarchy of interactions based on a layered structure where each layer is an ensemble of multiple heterogeneous interaction modules. Given the concatenation of $m$ input features $\X^{(l)} \in \mathbb{R}^{d \times m}$, the output of the $l$-th DHEN layer is computed as:
\begin{equation*}
    \X^{(l+1)} = \text{Norm}\left(\text{Ensemble}_{i=1}^k \text{Interaction}_i(\X^{(l)}) + \text{ShortCut}(\X^{(l)})\right),
\end{equation*}
where $\text{Norm}(\cdot)$ is a normalization function, $\text{Interaction}_i$ represents different interaction modules that are further assembled by $\text{Ensemble}_{i=1}^k(\cdot)$ such as summation and concatenation, and $\text{ShortCut}(\cdot)$ is an MLP serving as the residual connection for deep layer stacking.


\vspace{-2pt}
\subsection{Attention Mechanism}
Attention mechanisms~\cite{bahdanau2014neural} have become an integral part of sequence modeling, serving as the core component behind various model designs~\cite{vaswani2017attention,devlin2018bert}. We briefly review attention mechanisms, including Multi-Head Attention and Pooling by Multihead
Attention (PMA).

\noindent\textbf{Multi-Head Attention (MHA)~\cite{vaswani2017attention}} allows sequence modeling without regard to distance between tokens~\cite{vaswani2017attention}.
Given an input sequence $\S = [\s_1, \dots, \s_T]$, where $\s_t \in \mathbb{R}^d$ is the embedding vector at timestamp $t$, the self-attention operation is defined as:
\begin{equation}\label{eq:attn}
    \text{Attn}(\mathbf{Q}, \mathbf{K}, \mathbf{V}) = \text{softmax}\left(\mathbf{QK}^\T/\sqrt{d_k}\right)\mathbf{V},
\end{equation}
where $\mathbf{Q}, \mathbf{K}, \mathbf{V} \in \mathbb{R}^{n \times d_k}$ are query, key and value representations computed by linear projections $\mathbf{Q}=\S^\T\mathbf{W}^Q, \mathbf{K}=\S^\T\mathbf{W}^K, \mathbf{V}=\S^\T\mathbf{W}^V$.

To enable joint attention to information from different embedding subspaces, MHA aggregates $h$ parallel attention as follows:
\begin{equation}\label{eq:mha}
\text{MHA}(\mathbf{Q}, \mathbf{K}, \mathbf{V}) = \left[\textbf{head}_1\| \ldots\| \textbf{head}_h\right]\mathbf{W}^O,
\end{equation}
where heads are computed by self-attention $\textbf{head}_i = \text{Attn}(\mathbf{QW}_i^Q, \allowbreak \mathbf{KW}_i^K, \mathbf{VW}_i^V)$ and aggregated via an output projector $\mathbf{W}^O\in\mathbb{R}^{hd_k\times d}$.

\noindent\textbf{Pooling by Multi-Head Attention (PMA)~\cite{lee2019set}.}
Rather than obtaining query, key and values from the same input sequence $\S$, PMA utilizes a \textit{learnable} query $\mathbf{Q}_{\text{PMA}}\in\mathbb{R}^{k\times d_k}$ to summarize the sequence from $k$ different aspects, which can be defined as follows
\begin{equation}\label{eq:pma}
    \text{PMA}(\mathbf{Q}_{\text{PMA}}, \S)=\text{MHA}(\mathbf{Q}_{\text{PMA}}, \mathbf{K},\mathbf{V}).
\end{equation}
Intuitively, each column in $\mathbf{Q}_{\text{PMA}}$ is a seed vector summarizing $\S$ as a $d$-dimensional vector, and the output of PMA is the concatenation of $k$ summarization depicting sequence from different aspects.
\section{Methodology}\label{sec:method}
\begin{figure*}
    \centering
    \includegraphics[width=\linewidth]{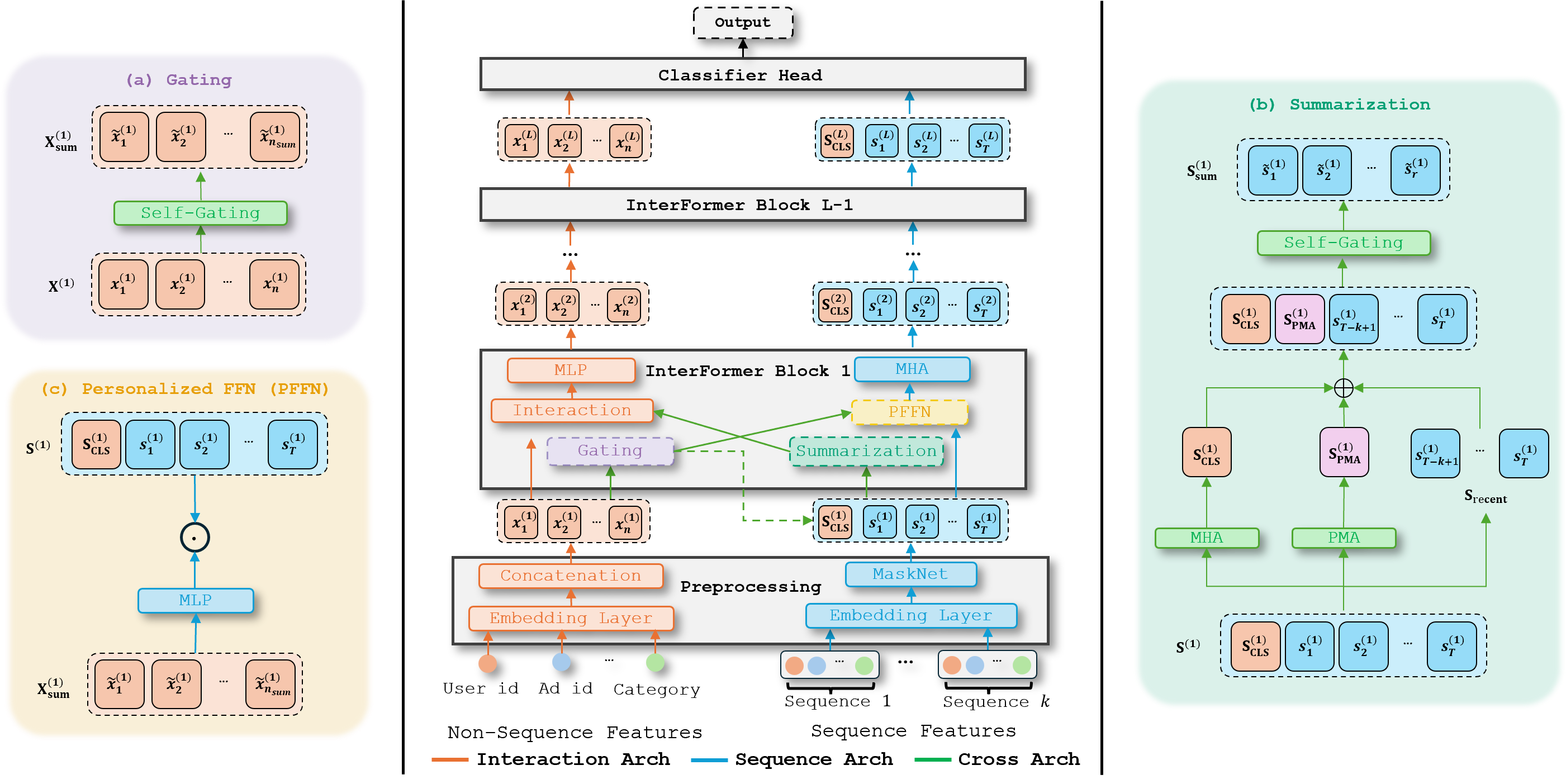}
    \vspace{-15pt}
    \caption{An overview of the \algname\ model architecture. Each block consists of three parts, including (1) Interaction Arch (orange) to learn behavior-aware non-sequence embeddings given sequence queries; (2) Sequence Arch (blue) to learn context-aware sequence embeddings given non-sequence queries; (3) Cross Arch (green) to exchange information between Interaction and Sequence Arch. Note that the dashed green line for CLS token appending only happens at the first layer.}
    \vspace{-10pt}
    \label{fig:interformer}
\end{figure*}

In this section, we present our proposed \algname.
We first introduce the preprocessing module in Section~\ref{sec:preproc}, followed by three major modules, including Interaction Arch, Sequence Arch and Cross Arch.
To learn behavior-aware non-sequence embeddings, the Interaction Arch models feature interactions (Section~\ref{sec:non-sequence}). 
To learn context-aware sequence embeddings, the Sequence Arch is proposed to optimize sequence (Section~\ref{sec:sequence}).
The Cross Arch connects Interaction and Sequence Arches, enabling effective information summarization and exchange between different modes (Section~\ref{sec:bridging}). An overview of \algname\ is shown in Figure~\ref{fig:interformer}.

\vspace{-2pt}
\subsection{Feature Preprocessing}\label{sec:preproc}
\noindent\textbf{Non-Sequence Feature Preproessing.}
Two types of non-sequence features are considered, including dense features like user age and item price, and sparse features like user id and item category.
To unify the heterogeneity in non-sequence features, different features are transformed into embeddings of the same dimensionality.

Specifically, raw dense features $x_{\text{dense}_i}^{(0)}$ are first concatenated to form a dense vector $\x_{\text{dense}}^{(0)}=\left[x_{\text{dense}_1}^{(0)},\dots,x_{\text{dense}_m}^{(0)}\right]^\T$, which is further transformed into a $d$-dimensional dense embedding vector via linear transformation, i.e., $\x_{\text{dense}}^{(1)}=\mathbf{W}_{\text{dense}}\x_{\text{dense}}^{(0)}$.
Similarly, each raw sparse feature is first encoded as an one-hot vector $\x_{\text{sparse}_i}^{(0)}\in\mathbb{R}^{n_{v_i}}$, where $n_{v_i}$ is the vocabulary size of the $i$-th sparse feature, and further transformed into a $d$-dimensional embedding vector by $\x_{\text{sparse}}^{(1)}=\mathbf{W}_{\text{sparse}}\x_{\text{sparse}}^{(0)}$.

By concatenating the dense and sparse embedding vectors, the input non-sequence embedding matrix can be obtained as follows
\begin{equation}\label{eq:non-sequencepreproc}
    \X^{(1)}=\left[\x_{\text{dense}}^{(1)}\|\x_{\text{sparse}_1}^{(1)}\|\dots\|\x_{\text{sparse}_n}^{(1)}\right].
\end{equation}

\noindent\textbf{Sequence Feature Preprocessing.}
Similar to the non-sequence feature preprocessing, an embedding layer is employed such that each interacted item in the sequence is mapped to a $d$-dimensional vector $\s_{t}$, and the behavior sequences is represented as the concatenation of item embeddings, i.e., $\S^{(0)}=\left[\s_1^{(0)}\|\dots\|\s_T^{(0)}\right]\in\mathbb{R}^{d\times T}$.

Real-world scenarios often encounter multiple sequences from different user actions (e.g., click and conversion)or various platforms.
Moreover, due to the uncertainty of user behaviors, these sequences often contain noisy and irrelevant user-item interactions.
To address this, a MaskNet~\cite{wang2021masknet} is employed to unify multiple sequences and filter out the internal noises via self-masking.

Specifically, given $k$ sequences $\S_1,\cdots,\S_k$, they are first concatenated along the embedding dimension, i.e., $\left[\S_1^{(0)^\T}\|\cdots\|\S_k^{(0)^\T}\right]^\T\in\mathbb{R}^{kd\times T}$, and further processed by the MaskNet operation as follows
\begin{equation}\label{eq:seqpreproc}
    \text{MaskNet}(\S)=\text{MLP}_{\text{lce}}\left(\S\odot\text{MLP}_{\text{mask}}(\S)\right),
\end{equation}
where $\text{MLP}_{\text{mask}}\!: \mathbb{R}^{kd\times T}\!\!\to\! \mathbb{R}^{kd\times T}$ computes self-masking to select relevant information from input sequences, and $\text{MLP}_{\text{lce}}: \mathbb{R}^{kd\times T}\!\to\! \mathbb{R}^{d\times T}$ linearly combines multiple sequences into one, matching the dimensionalities of non-sequence and sequence features.

\vspace{-2pt}
\subsection{Interaction Arch: Towards Behavior-Aware Interaction Learning}\label{sec:non-sequence}

Non-sequence features, such as user profile and ad content, provide substantial information in understanding user preference over a specific item.
Modeling the interaction among non-sequence features is the key towards the success of CTR prediction~\cite{rendle2010factorization, wang2021dcn}.
While non-sequence features reflect static user interests~\cite{lian2018xdeepfm,zhang2022dhen}, the behavior sequence provides complementary information depicting user interests from a dynamic view~\cite{zhou2019deep}. 
For instance, while user profile exhibits a general interest in electronics, the recent browsing history on smartphones offers more specific and timely information on current needs.
Therefore, it is crucial to learn behavior-aware non-sequence interactions to adapt to the evolving environment.

To learn behavior-aware non-sequence interaction, we model the interactions among non-sequence features, as well as sequence summarization, by an interaction module.
Formally, given the non-sequence input $\X^{(l)}$ and sequence summarization $\S_\text{sum}^{(l)}$ at the $l$-th layer, the output of $l$-th Interaction Arch is defined as:
\begin{equation}\label{eq:non-sequence}
    \X^{(l+1)}=\text{MLP}^{(l)}\left(\text{Interaction}^{(l)}\left(\left[\X^{(l)}\|\S_{\text{sum}}^{(l)}\right]\right)\right),
\end{equation}
where $\text{Interaction}^{(l)}(\cdot)$ is the interaction module.
Note that we do not adhere to a specific interaction module, but rather, various backbone models, such as inner product, DCNv2~\cite{wang2021dcn} and DHEN~\cite{zhang2022dhen}, can be adopted.
Additionally, an MLP is used to transform the output $\X^{(l+1)}$ to match the shape of the input $\X^{(l)}$ enabling selective information aggregation and facilitating easy layer stacking.

By performing interaction learning on the concatenation of $\X^{(l)}$ and $\S^{(l)}_{\text{sum}}$, Eq.~\eqref{eq:non-sequence} incorporates non-sequence and sequence features.
The interactions among non-sequence features capture explicit user interests by computing the relevance between user profile and target item content, while the interactions between non-sequence and sequence capture implicit user interests by computing the relevance between target item and user's current need within the interaction sequence.
Additionally, by computing the interactions among sequence features, the most representative information is promoted with high interaction scores, while inactive items with low scores are filtered out. By stacking multiple layers, the Interaction Arch captures rich behavior-aware interactions at different orders.

\vspace{-2pt}
\subsection{Sequence Arch: Towards Context-Aware Sequence Modeling}\label{sec:sequence}
In addition to the explicit static user preference in non-sequence features, the implicit dynamic user interests behind behavior sequences provide complementary information~\cite{zhou2019deep, sun2019bert4rec}.
However, given that behavior sequences are highly random and noisy, solely relying on sequential information is ineffective.
It is important to incorporate non-sequential context into sequence modeling.
For example, a user may randomly browse items on shopping platforms, but given the static information that the user is an electronic fan, electronic items, e.g., smartphones and laptops, in the browsing history provide key information demanding extra attention.

To learn context-aware sequence embeddings, a Sequence Arch is designed based on two key ideas: Personalized FFN (PFFN) and Multi-Head Attention (MHA). To enable interactions between non-sequential and sequential information, PFFN is employed to transform sequence embeddings given non-sequence summarization as query. Given non-sequence summarization $\X_\text{sum}^{(l)}$ and sequence embedding $\S^{(l)}$ at the $l$-th layer, the PFFN operation is defined as:
\begin{equation}
    \text{PFFN}\left(\X_\text{sum}^{(l)}, \S^{(l)}\right)=f\left(\X_\text{sum}^{(l)}\right)\S^{(l)},
    \vspace{-5pt}
\end{equation}
where $f\left(\X_\text{sum}^{(l)}\right)$ is an MLP that aims to learn the linear projection on sequence based on non-sequence summarization.

Besides, to model the relationship among events in a sequence, MHA is applied to enable the model to attend to different parts of the sequence, capturing long-range dependencies and contextual information.
To incorporate non-sequential context, before feeding into the first \algname\ layer, the non-sequence summarization $\X^{(1)}_\text{sum}$ is prepended to the sequence as the CLS token, i.e., $\S^{(1)}=\left[\X^{(1)}_\text{sum}\|\S^{(1)}\right]$, hence the following MHA on the CLS token can aggregate sequential information using non-sequential information as the query~\cite{vaswani2017attention,devlin2018bert}. This is similar to the early fusion idea described in Transact ~\cite{xia2023transact} but we primarily consider 'append' instead of 'concat' as the fusion method. 
Additionally, rotary position embeddings~\cite{su2024roformer} are applied on tokens so that the positional information in sequences can be effectively leveraged.
In general, the Sequence Arch can be written as follow:
\begin{equation}\label{eq:sequence}
    \S^{(l+1)}=\text{MHA}^{(l)}\left(\text{PFFN}\left(\X_\text{sum}^{(l)}, \S^{(l)}\right)\right).
    \vspace{-5pt}
\end{equation}
As the output $\S^{(l+1)}$ is of the same shape as the input $\S^{(l)}$, aggressive aggregation can be avoided and layers can be easily stacked.

Through layer stacking, for one thing, the sequence embeddings is aware of non-sequence interactions at different orders via PFFN; for another, MHA at different layers focuses on different parts of the sequence, capturing multi-scale sequential information. Therefore, the model can learn a rich context-aware encoding of the sequential data capturing both local and global patterns within the sequence.

\vspace{-2pt}
\subsection{Cross Arch: Towards Effective Information Selection and Summarization}\label{sec:bridging}
Though selective information aggregation is achieved in both the Interaction and Sequence Arch, as the dimensionalities of input and output features are retained till the final layer, it is infeasible to directly exchange such information due to (1) \textit{ineffectiveness} given the noisy information and (2) \textit{inefficiency} given the high-dimensionality.
To solve this dilemma, we propose a cross Arch to select and summarize information before exchanging information.

To start with, given the large scale of non-sequence embeddings, it is important to selectively summarize them to guide sequence learning. To this end, we highlight the most useful information through a personalized gated selection mechanism as follows:
\begin{equation}\label{eq:non-sequence_sum}
    \begin{aligned}
        &\X_{\text{sum}}^{(l)}=\text{Gating}(\text{MLP}(\X^{(l)})),\\
        &\text{where }\text{Gating}(\X)=\sigma\left(\X\odot \text{MLP}(\X)\right).
    \end{aligned}
\end{equation}
where MLP: $\mathbb{R}^{d\times n}\to \mathbb{R}^{d\times n_{\text{sum}}}$ with $n_{\text{sum}}\ll n$, and $\sigma$ is the activation function, e.g., sigmoid, tanh or even identity functions. Self-gating \cite{chai2020highway} provides sparse masking on the embeddings such that relevant information is retained while irrelevant noises are filtered out, providing high-quality context for sequence learning. 

For sequential information, three types of summarization are neatly designed, including CLS tokens, PMA tokens and recent interacted tokens.
The CLS tokens $\S_{\text{CLS}}$ learned by MHA are selected as context-aware sequence summarization.
However, the quality of CLS tokens largely depend on the learned non-sequential context.
To compensate such heavy reliance and enable more flexibility, the PMA tokens $\S_{\text{PMA}}$~\cite{lee2019set}, which are essentially sequence summarization based on learnable queries, are employed. 
Besides, the $K$ most recent interacted tokens ($\S_{\text{recent}}$) have been proven to be effective in capturing the user's recent interests~\cite{borisyuk2024lirank, xia2023transact}.
The combination of the above information is further gated by a self-gating layer, serving as the behavior summarization for non-sequence interaction:
\begin{equation}\label{eq:sequence_sum}    \S^{(l)}_{\text{sum}}=\text{Gating}\left(\left[\S^{(l)}_{\text{CLS}}\|\S^{(l)}_{\text{PMA}}\|\S^{(l)}_{\text{recent}}\right]\right).
\end{equation}

In general, the benefits of the cross Arch are two-fold.
On the one hand, by separating information summarization from Interaction and Sequence Arch, information can be retained in both arches, avoiding aggressive information aggregation.
On the other hand, effective information exchange can be achieved as high-dimensional non-sequence/sequence features are selected and summarized into low-dimensional embeddings.
Therefore, the Cross Arch plays a pivotal role in enabling the model to capture complex interactions between non-sequence behaviors and sequential patterns, leading to more comprehensive representations of the input data.

\section{Experiment}\label{sec:exp}
In this section, we carry out extensive experiments to evaluate the proposed \algname.
We first introduce the experiment setup in Section~\ref{sec:exp-setup}.
Afterwards, we evaluate \algname\ on public benchmark datasets are carried out in Section~\ref{sec:exp-benchmark}.
The proposed \algname\ is deployed at Meta, and the post-launch results on internal industrial-scale dataset are provided in Section~\ref{sec:exp-industrial}.
\vspace{-5pt}
\subsection{Experiment Setup}\label{sec:exp-setup}
\noindent\textbf{Datasets.} 
We adopt three benchmark datasets, including AmazonElectronics~\cite{he2016ups}, TaobaoAd~\cite{Tianchi}, KuaiVideo~\cite{li2019routing}, and a large-scale internal dataset for evaluation. Dataset statistics are summarized in Table~\ref{tab:data} with more details in Appendix~\ref{app:data}.
\vspace{-5pt}
\begin{table}[htbp]
\caption{Dataset Summary.}
\vspace{-10pt}
\label{tab:data}
\begin{tabular}{@{}llll@{}}
\toprule
Dataset           & \#Samples & \#Feat. (Seq/Non-Seq) & Seq Length \\ \midrule
Amazon            & 3.0M      & 6 (2/4)               & 100                            \\
TaobaoAd          & 25.0M     & 22 (3/19)             & 50                             \\
KuaiVideo         & 13.7M     & 9 (4/5)               & 100                            \\ \bottomrule
\end{tabular}
\vspace{-5pt}
\end{table}

\noindent\textbf{Baseline Methods.} We compare the proposed \algname\ with 11 state-of-the-art models, including (1) non-sequential methods: FM~\cite{rendle2010factorization}, xDeepFM~\cite{lian2018xdeepfm}, AutoInt+~\cite{song2019autoint}, DCNv2~\cite{wang2021dcn}, FmFM~\cite{sun2021fm2}, DOT product, DHEN~\cite{zhang2022dhen}, Wukong~\cite{zhang2024wukong}, and (2) sequential methods: DIN~\cite{zhou2018deep}, DIEN~\cite{zhou2019deep}, BST~\cite{chen2019behavior}, DMIN~\cite{xiao2020deep}, DMR~\cite{lyu2020deep}, TransAct~\cite{xia2023transact}. We adopt DHEN as the Interaction Arch in the experiments.
Detailed model configurations are provided in Appendix~\ref{app:config}.

\noindent\textbf{Metrics.} We adopt four widely-used metrics to evaluate the models from different aspects, including:
\begin{itemize}
    \item \texttt{AUC} provides an aggregated measure of model capacity in correctly classifying positive and negative samples across all thresholds. Higher the better.
    \item \texttt{gAUC} provides personalized AUC evaluation, where users are weighted by their click account. Higher the better.
    \item \texttt{LogLoss} (cross-entropy loss) measures the distance between the prediction $\hat{y}$ and the label $y$, and is computed as $L(y,\hat{y})\!=\!-\left(y\log\left(\hat{y}\right)+(1\!-\!y)\log\left(1\!-\!\hat{y}\right)\right)$. Lower the better.
    \item \texttt{NE} (normalized entropy)~\cite{he2014practical}, is the \texttt{LogLoss} normalized by the entropy of the average empirical CTR of the training set. NE  provides a data-insensitive evaluation on model performance as the normalization alleviates the effect of background CTR on model evaluation. Lower the better.
\end{itemize}

\subsection{Evaluation on Benchmark Datasets}\label{sec:exp-benchmark}

\subsubsection{Results.} The experiment results are shown in Table~\ref{tab:exp-result}. We first observe that sequential methods consistently outperforms non-sequential methods on all datasets.
For non-sequential methods, sequence information is naively aggregated in early stages and further processed together with non-sequence information.
While in sequential methods, neatly designed sequence processing modules, e.g., RNN~\cite{zhou2019deep, sun2019bert4rec} and attention mechanism~\cite{zhou2018deep,chen2019behavior,xiao2017attentional,lyu2020deep}, are employed, so that sequential information can be processed in aware of non-sequence context.
The universal outperformance of sequential methods validates that different data modes should be processed differently.
Besides, results show that aggressive sequence summarization in the early stages, without considering non-sequence context, will impair model performance.

Comparing \algname\ with other sequential methods, \algname\ achieves state-of-the-art performance. Specifically, \algname\ outperforms the best competitor by up to 0.9\% in gAUC, 0.14\% in AUC and 0.54\% in LogLoss. These results demonstrate \algname's effectiveness on diverse datasets and generalization across different recommendation tasks.

\begin{table*}[htbp]
\caption{Experiments on benchmark datasets. Methods with high gAUC and AUC, and low LogLoss and \#Params are preferred.}
\vspace{-5pt}
\label{tab:exp-result}
\setlength\tabcolsep{4pt}
\begin{tabular}{l|llll|llll|llll}
\toprule
\textbf{Dataset} & \multicolumn{4}{c|}{\textbf{AmazonElectronics}}       & \multicolumn{4}{c|}{\textbf{TaobaoAds}}             & \multicolumn{4}{c}{\textbf{KuaiVideo}}          \\
\textbf{Metric} & \textbf{gAUC}   & \textbf{AUC}     & \textbf{LogLoss} & \textbf{\#Params} & \textbf{gAUC}  & \textbf{AUC}  & \textbf{LogLoss}  & \textbf{\#Params} & \textbf{gAUC}  & \textbf{AUC}  & \textbf{LogLoss}  & \textbf{\#Params} \\ \midrule
FM                               & 0.8494          & 0.8485          & 0.5060  & 4.23M        & 0.5628        & 0.6231       & 0.1973 & 43.08M         & 0.6567        & 0.7389       & 0.4445    & 52.76M\\
xDeepFM                          & 0.8763          & 0.8791          & 0.4394      & 5.46M    & 0.5675        & 0.6378       & 0.1960  & 53.79M         & 0.6621        & 0.7423       & 0.4382           & 43.56M     \\
AutoInt+                         & 0.8786          & 0.8804          & 0.4441      & 5.87M    & 0.5701        & 0.6467       & 0.1941   & 42.05M       & 0.6619              & 0.7420             & 0.4369                 & 43.95M         \\
DCNv2                            & 0.8783          & 0.8807          & 0.4447      & 5.23M    & 0.5704        & 0.6472       & 0.1933   & 43.71M       & 0.6627        & 0.7426       & 0.4378           & 42.48M    \\
FmFM                             & 0.8521          & 0.8537          & 0.4796    & 4.21M       & 0.5698        & 0.6330       & 0.1963  & 43.06M         & 0.6552               & 0.7389             & 0.4429                 & 51.97M      \\
DOT                             & 0.8697          & 0.8703          & 0.4485    & 4.23M      & 0.5701        & 0.6482       & 0.1941    & 41.54M       & 0.6605              & 0.7435             & 0.4361    & 41.29M               \\
DHEN                             & 0.8759          & 0.8790          & 0.4398      & 4.99M    & 0.5708      & 0.6509       & 0.1929  & 43.89M        & 0.6589     & 0.7424       & 0.4365       & 42.06M \\
Wukong       & 0.8747            & 0.8765          & 0.4455  & 4.28M        & 0.5693           & 0.6478        & 0.1932   &41.72M    & 0.6587           & 0.7423         & 0.4372      &41.37M \\ \midrule
DIN                              & 0.8817          & 0.8848          & 0.4324    & 5.40M      & 0.5719        & 0.6507       & 0.1931  & 42.26M        & 0.6621        & 0.7437       & 0.4353           & 43.03M    \\
DIEN                             & 0.8825          & {\ul 0.8856}          & {\ul 0.4276}     & 5.37M     & 0.5721        & {\ul 0.6519}       & {\ul 0.1929}      & 42.38M    & \textbf{0.6651}              & {\ul 0.7451}             & {\ul 0.4343}                 & 43.43M      \\
BST                              & 0.8823          & 0.8847          & 0.4305      & 5.30M     & 0.5698        & 0.6489       & 0.1935      & 42.05M    & 0.6617              & 0.7446              & 0.4352                 & 42.83M          \\
DMIN                             & 0.8831          & 0.8852          & 0.4298     & 5.94M   & {\ul 0.5723}        & 0.6498       & 0.1933       & 42.17M    & 0.6623              & 0.7449             & 0.4356                 & 41.61M    \\
DMR                              & 0.8827          & 0.8848          & 0.4309     & 6.47M     & 0.5711        & 0.6504       & 0.1932    & 45.82M       & {\ul 0.6642}              & 0.7449             & 0.4355                & 44.15M   \\ 
TransAct & {\ul 0.8835} & 0.8851 & 0.4285 & 7.56M & 0.5715 & 0.6498 & 0.1933 & 44.39M & 0.6632 & 0.7448 & 0.4352 & 42.97M \\ \midrule
\algname                             & \textbf{0.8843}    & \textbf{0.8865} & \textbf{0.4253}   & 7.18M   & \textbf{0.5728}        & \textbf{0.6528}       & \textbf{0.1926}      &  44.73M     & 0.6637        & \textbf{0.7453}       & \textbf{0.4340}     & 43.61M    \\ \bottomrule
\end{tabular}
\end{table*}

\subsubsection{Analysis of the Model.}
\begin{figure*}[htbp]
    \centering
    \begin{subfigure}[b]{0.33\textwidth}
        \centering
        \includegraphics[width=\textwidth, trim=10 0 10 0, clip]{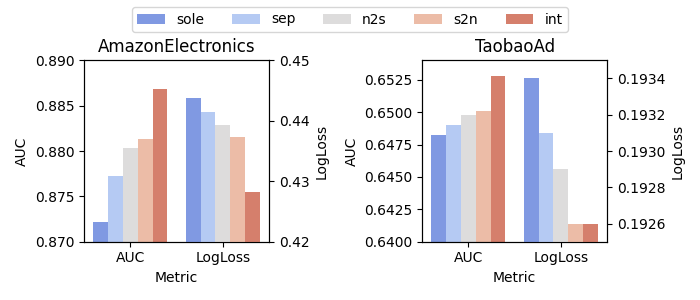}
        \vspace{-15pt}
        \caption{DOT}
    \end{subfigure}
    \begin{subfigure}[b]{0.33\textwidth}
        \centering
        \includegraphics[width=\textwidth, trim=10 0 10 0, clip]{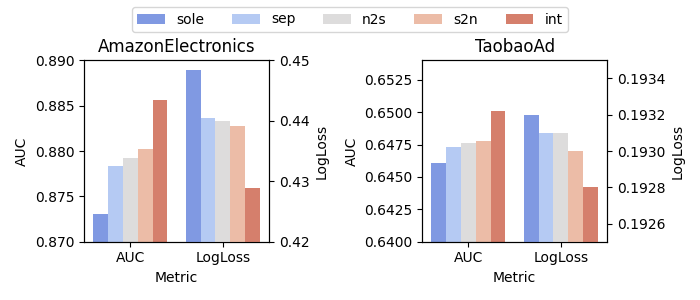}
        \vspace{-15pt}
        \caption{DCNv2}
    \end{subfigure}
    \begin{subfigure}[b]{0.33\textwidth}
        \centering
        \includegraphics[width=\textwidth, trim=10 0 10 0, clip]{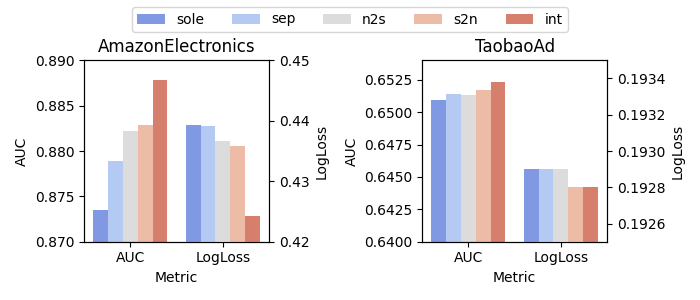}
        \vspace{-15pt}
        \caption{DHEN}
    \end{subfigure}
    \vspace{-20pt}
    \caption{Study on the interleaving learning style. We consider different non-sequence backbones (DOT, DCNv2, DHEN) and different scenarios (\texttt{sole, sep, n2s, s2n, int}). With more information exchanged between different data modes, we observe an ascending order in performance: \texttt{sole}<\texttt{sep}<\texttt{n2s}$\approx$\texttt{s2n}<\texttt{int}, validating the necessity of the interleaving learning style (\texttt{int}).}
    \label{fig:backbone}
\end{figure*}
First (\underline{\textit{Interleaving Learning Style}}), we analyze how the interleaving learning style benefits heterogeneous interaction learning. To validate the universality of \algname, we consider three backbone Interaction Arch, including dot product (\texttt{DOT}), \texttt{DCNv2}, and \texttt{DHEN}, and five different scenarios, including
\begin{itemize}
    \item \texttt{sole} where only Interaction Arch is adopted, while sequence information is naively aggregated in early stage.
    \item \texttt{sep} where inter-mode interaction is disabled and Sequence and Interaction Arch are learnt separately.
    \item \texttt{s2n} where only sequence-to-non-sequence information flow is enabled, while the reverse direction is disabled.
    \item \texttt{n2s} where only non-sequence-to-sequence information flow is enabled, while the reverse direction is disabled.
    \item \texttt{int} where the bidirectional information flows are activated.
\end{itemize}
As shown in Figure~\ref{fig:backbone}, \texttt{int} consistently outperforms other scenarios regardless of the backbone Interaction Arch, with an up to 1.46\% outperformance in AUC, validating the universal benefits brought by our proposed interleaving learning style. We also observe an ascending order in model performance when more information is exchanged between different data modes, i.e., \texttt{sole}<\texttt{sep}<\texttt{n2s}$\approx$\texttt{s2n}<\texttt{int}. 
This validates our claim that the insufficient inter-mode interaction is a key bottleneck of heterogeneous interaction learning, and bidirectional information flow enables different data modes to be learnt in a mutually beneficial manner. 
Specifically, when equipped with an additional Sequence Arch (\texttt{sep}), the model performance consistently outperforms \texttt{sole} w/o sequence modeling.
This not only shows the necessity of employing different arch for different data modes, but also implies the possible performance degradation when heterogeneous data is integrated naively. 
Furthermore, given that the bidirectional information flow (\texttt{int}) consistently outperforms the unidirectional setting (\texttt{n2s} and \texttt{s2n}), we attribute the outperformance of \algname\ on other state-of-the-art sequential methods~\cite{zhou2019deep,xiao2020deep,chen2019behavior} to the bidirectional information flow.

\begin{figure}[t]
    \centering
    \includegraphics[width=\linewidth, clip, trim=0 12 0 5]{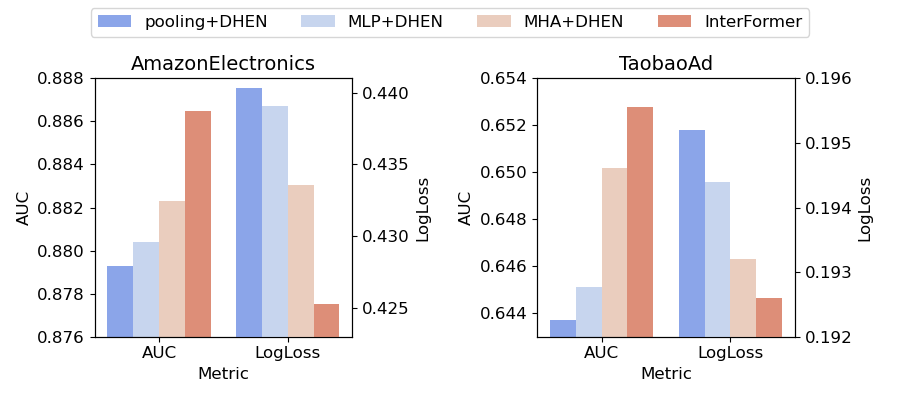}
    \vspace{-15pt}
    \caption{Study on information aggregation. We compare the performance with early summarization via average pooling, MLP, and MHA, and \algname's selective aggregation.}
    \vspace{-10pt}
    \label{fig:aggregation}
\end{figure}

Second (\underline{\textit{Selective Information Aggregation}}), we evaluate the effect of selective information aggregation by comparing with three aggressive aggregation variants, where sequence information is first compressed by (1) \texttt{average pooling}, (2) \texttt{MLP}, and (3) \texttt{MHA}, and further combined with non-sequence features by the interaction module DHEN. The experimental results are shown in Figure~\ref{fig:aggregation}.
As the results suggest, selective aggregation (\algname) consistently achieves the best performance compared with other aggressive variants. In general, we can regard MLP as a more selective version of average pooling as MLP gradually compresses the sequence information, and similarly, MHA process sequence less aggressive than MLP. Therefore, we may conclude that with more selective aggregation, the CTR prediction quality improves.

\begin{figure}[t]
    \centering
    \vspace{-10pt}
    \includegraphics[width=\linewidth, clip, trim=5 5 5 10]{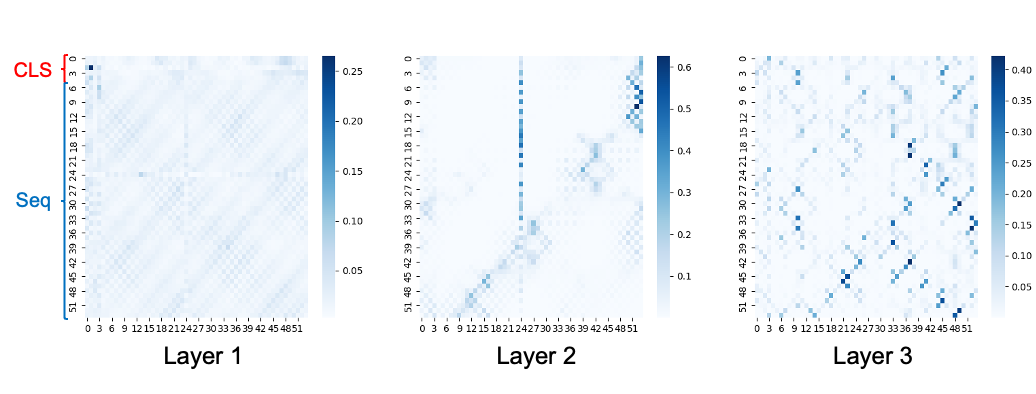}
    \vspace{-20pt}
    \caption{Attention map on TaobaoAds. The first 4 tokens are CLS tokens, followed by the behavior sequence of length 50.}
    \vspace{-10pt}
    \label{fig:attn}
\end{figure}

Third (\underline{\textit{Sequence Modeling}}), we visualize the learned attention map in Figure~\ref{fig:attn} to understand sequence modeling.
We observe that the attention map exhibits a cluster structure, i.e., surrounding tokens are likely to interact, which can be regarded as a weighted pooling within selected surrounding tokens. Such selective pooling process alleviates random noises in the behavior history to extract better temporal user interests. Besides, MHAs at different layers focus on different tokens at different scales. For example, the attention map is quite uniform in the first layer, indicating pooling in a broad scale for long-term interests. In contrast, the third layer exhibits stronger interactions within smaller clusters, resulting in fine-grained aggregation for short-term interests. Furthermore, the second layer focuses on specific tokens, e.g., recent tokens (tokens in right columns) and the 24-th token, for item-specific information.
\vspace{-10pt}

\subsubsection{Ablation Study}
\begin{figure}[t]
    \centering
    \includegraphics[width=\linewidth, clip, trim=0 12 0 5]{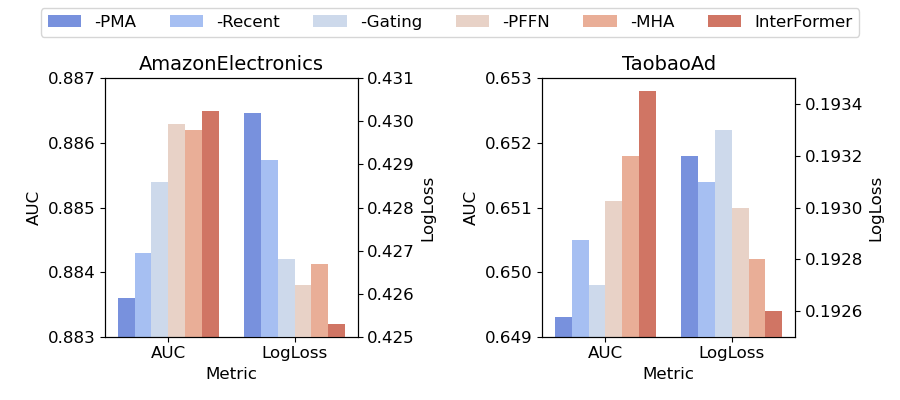}
    \vspace{-15pt}
    \caption{Ablation study on \algname, where '-' indicates ablating different information or modules.}
    \vspace{-10pt}
    \label{fig:ablatation}
\end{figure}
To better understand how different model designs contribute to the final performance, we ablate different exchanged information, including PMA tokens and recent interacted tokens, as well as different modules like the gating module, PFFN and MHA. As the results shown in Figure~\ref{fig:ablatation}, all the exchanged information and modules contribute to the model performance. Specifically, PMA tokens provide effective sequence summarization given non-sequence context and are of great importance, ablating which decreases AUC by up to 0.004. The gating modules help selective summarize information from different data modes contribute up to 0.003 AUC improvement. The sequence modeling modules (PFFN and MHA) also improves the performance to some extent.

\vspace{-5pt}
\subsection{Evaluation on Industrial Datasets}\label{sec:exp-industrial}
To evaluate \algname's performance at industrial-scale data, we carry out experiments on a large-scale internal dataset from Meta, containing 70B samples in total, hundreds of non-sequence features and 10 sequences of length 200 to 1,000. 

\subsubsection{Results.}
In general, a 3-layer \algname\ achieves a 0.15\% NE gain compared to the internal SOTA model with similar FLOPs and 24\% Queries Per Second (QPS) gain. Together with feature scaling, the improvement on NE can be further enlarged with a 10\% of Model FLOPs Utilization (MFU) on 512 GPUs attained. \algname\ shows great generalizability on a wide range of models showing promising ROI.

\begin{figure}
    \centering
    \begin{subfigure}[b]{0.48\linewidth}
        \centering
        \includegraphics[width=\textwidth]{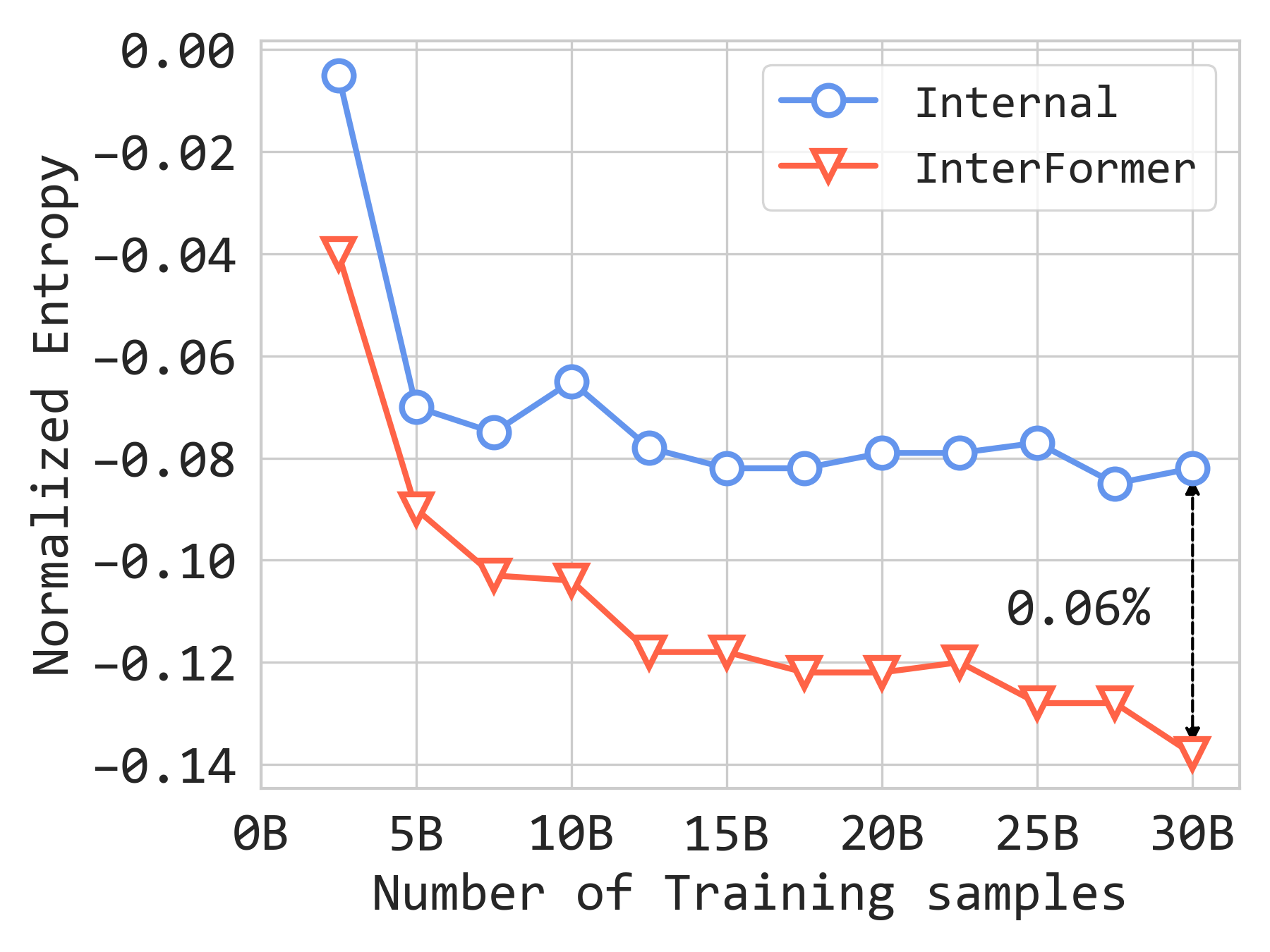}
        \caption{Feature scaling}
        \label{fig:feat_scale}
    \end{subfigure}
    \begin{subfigure}[b]{0.48\linewidth}
        \centering
        \includegraphics[width=\textwidth]{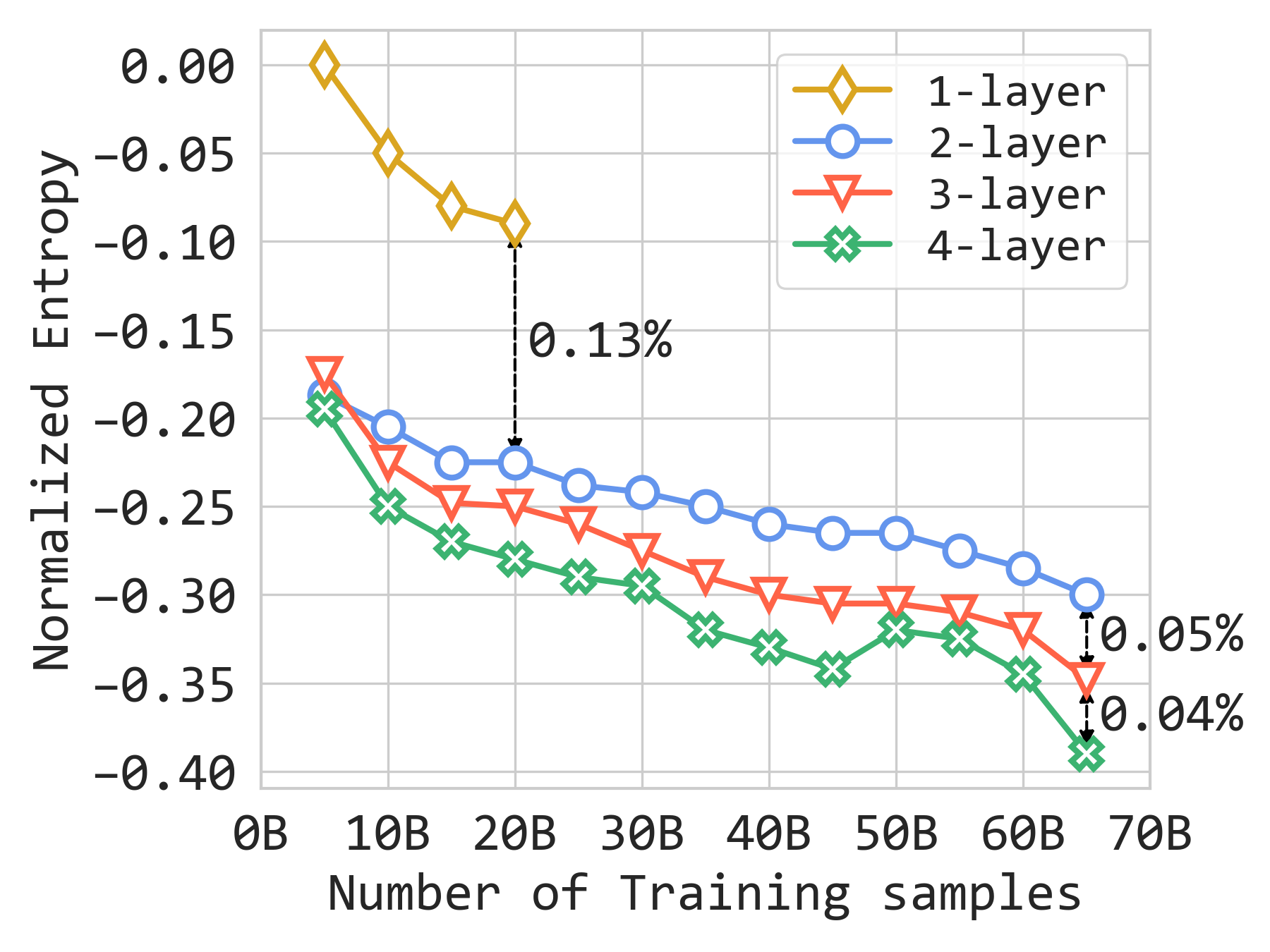}
        \caption{Model scaling}
        \label{fig:model_scale}
    \end{subfigure}
    \vspace{-5pt}
    \caption{Scalability results on (a) feature scaling and (b) model scaling. \algname\ achieves better scaling curve than \textsc{Internal}, and promising layer-stacking performance.}
    \vspace{-10pt}
    \label{fig:scale}
\end{figure}

Besides (\underline{\textit{Sequence Feature Scaling}}), we evaluate how the performance of \algname\ changes when the sequence feature scale increases in internal dataset. In addition to six sequences of length 100, we include two additional long sequences of length 1000, and we observe a 0.14\% improvement in NE. As shown in Figure~\ref{fig:feat_scale}, \algname\ exhibits better scalability compared to the strong internal baseline (\textsc{Internal}) that leverages cross-attention to capture sequence and non-sequence information, as NE curve of \algname\ continues to decrease when more training samples are involved, outperforming \textsc{Internal} by 0.06\% in NE. Besides, we also tried to merge six sequences to generate one long sequence of length 600, and we observe promising efficiency improvements in QPS (+20\%) and MFU (+17\%) with a 0.02\% NE tradeoff. From the modeling perspective, the results validate that \algname\ is able to enlarge the NE gain brought by the sequence feature scaling.

In addition (\underline{\textit{Model Scaling}}), we evaluate how \algname\ performs when the model scale increases. As Figure~\ref{fig:model_scale} shows, scaling \algname\ from 1 to 4 layers achieves consistent NE gains, exhibiting good scaling properties. Specifically, compared to a single layer \algname, a two-layer \algname\ achieves a significant 0.13\% NE gain, with an additional 0.05\% and 0.04\% NE gain achieved by stacking 3 and 4 layers, respectively.

\subsubsection{Model-System Co-Design.}
Model architecture design plays a crucial role in training efficiency with its implication on GPU FLOPs utilization and inter-GPU communication. We highlight two optimizations based on \algname\ architecture that in total boost training efficiency by more than 30\%, including \textit{communication overhead reduction} and \textit{computation efficiency}.

For \textit{communication overhead reduction}, the Interaction Arch DHEN with heavy parameters and relatively light computation tends to be FSDP~\cite{zhao2023pytorch} communication-bound in distributed training, while the Sequence Arch Transformer with much higher computation (FLOPs) to parameter ratio is normally computation-bound.
To alleviate such inefficiency, our parallel design of Interaction and Sequence Arch allows the exposed communication from DHEN modules to overlap with sequence computation effectively, resulting in a 20\% QPS improvement compared to the performing two modules sequentially.

For \textit{computation efficiency}, we perform a series of optimizations, including 1) reallocating FLOPs from small, low-return modules to larger, high ROI modules and 2) combining smaller kernels to better utilize GPU resources. These optimizations improve MFU for interaction modules from 11\% to 16\%, and DHEN from 38\% to 45\%, with a 19\% MFU improvement for the overall \algname\ layer.

\subsubsection{Online Impact}
Interformer has been deployed in key Ads models, including the largest ones through \cite{luom3c}, showing superior ROI in scaling compared to other state-of-the-art architectures. Pilot launches in 2024 have yielded a 0.6\% improvement in topline metrics. We expect great impact to be delivered in subsequent deployment.
\section{Conclusion}\label{sec:con}
In this paper, we propose a novel building block for heterogeneous interaction learning in CTR prediction named \algname. The key idea is an interleaving learning style between different data modes to achieve effective inter-mode interaction, as well as selective information aggregation. In general, An Interaction Arch and a Sequence Arch are adopted to achieve behavior-aware interaction learning and context-aware sequence modeling, respectively. A Cross Arch is further proposed to for effective information selection and summarization. \algname\ achieves consistent outperformance on benchmark datasets, and 0.15\% improvement on the NE of CTR prediction on internal large-scale dataset. Pilot launches in key Ads models have yielded a 0.6\% improvement in topline metrics and we expect great impact to be delivered.

\newpage
\bibliographystyle{ACM-Reference-Format}
\balance
\bibliography{main}

%
\newpage
\appendix
\section*{Appendix}
\section{Model Design}
\subsection{Algorithm}
We present the overall algorithm of \algname\ in Algorithm~\ref{algo:interformer}.
\begin{algorithm}[h]
    \caption{\algname}\label{algo:interformer}
    \begin{algorithmic}[1]
        \REQUIRE Non-sequence feature $\X^{(0)}$, sequence feature $\S^{(0)}$, number of layers $L$;
        \ENSURE CTR prediction $\hat{y}$;
        \STATE Preprocess $\X^{(0)},\S^{(0)}$ via Eqs.~\eqref{eq:non-sequencepreproc} and \eqref{eq:seqpreproc} to obtain $\X^{(1)},\S^{(1)}$;
        \STATE Compute non-sequence summarization $\X_{\text{sum}}^{(1)}$ via Eq.~\eqref{eq:non-sequence_sum};
        \STATE Prepend $\X_{\text{sum}}^{(1)}$ before $\S^{(1)}$ as CLS tokens;
        \FOR{$l=1,2,\dots,L$}
            \STATE \texttt{Cross Arch}: compute non-sequence summarization $\X^{(l)}_{\text{sum}}$ and sequence summarization $\S^{(l)}_{\text{sum}}$ by Eqs.~\eqref{eq:non-sequence_sum} and \eqref{eq:sequence_sum};
            \STATE \texttt{Interaction Arch}: compute non-sequence embeddings $\X^{(l+1)}$ given $\X^{(l)}$ and $\S^{(l)}_{\text{sum}}$ by Eq.~\eqref{eq:non-sequence}; 
            \STATE \texttt{Sequence Arch}: compute sequence embeddings $\S^{(l+1)}$ given $\S^{(l)}$ and $\X^{(l)}_{\text{sum}}$ by Eq.~\eqref{eq:sequence};
        \ENDFOR
        \STATE Compute CTR prediction by $\hat{y}=\text{MLP}\left(\left[\X^{(L)}_{\text{sum}}\|\S^{(L)}_{\text{sum}}\right]\right)$;
        \RETURN CTR prediction $\hat{y}$.
    \end{algorithmic}
\end{algorithm}

\subsection{More Details on Module Blocks}
We provide more details on module blocks involved in \algname\ architecture, including Linear Compressed Embedding (LCE), and personalized FeedForward Network (PFFN).
\subsubsection{Linear Compressed Embedding} 
Given numerous non-sequence features, it is essential to compress embeddings to a manageable size. In general, given $N$ $d$-dimensional features $\mathbf{X}\in\mathbb{R}^{d\times N}$, LCE is a linear transformation on the sample dimension $\mathbf{W}\in\mathbb{R}^{N\times M}$, such that the linear transformation $\mathbf{XW}\in\mathbb{R}^{d\times M}$ serves as the compressed embedding with $M$ features. Together with self-gating modules, LCE selectively compress and denoise numerous embeddings, benefiting the model in both efficiency and effectiveness.

\subsubsection{Personalized FFN}
PFFN learns interactions between non-sequence and sequence embeddings, whose key idea is to apply an FFN on the sequence embeddings with weight learned based on the summarized non-sequence embeddings.
Given non-sequence summarization $\mathbf{X}_{\text{sum}}\in\mathbb{R}^{d\times n_{\text{sum}}}$ and sequence embedding $\mathbf{S}\in\mathbb{R}^{d\times T}$, PFFN first learns the transformation weight $\mathbf{W}_{\text{PFFN}}=\mathbf{X}_{\text{sum}}\mathbf{W}\in\mathbb{R}^{d\times d}$, which is further applied on the feature dimension of sequence embeddings, i.e., $\mathbf{W}_{\text{PFFN}}\mathbf{S}$.
In contrast to the vanilla FFN in Transformer that may not add too much value to subsequent sequence modeling, PFFN integrate side information from other data modes. In fact, PFFN is essentially computing enhanced dot-products based interaction between sequence embeddings and summarized non-sequence embeddings that brings much more value for the following sequence modeling.

\section{Experiment Pipeline}\label{app:pipeline}
We adopt the public BARS evaluation framework~\cite{zhu2022bars} for benchmark experiments.
In general, an Adam~\cite{kingma2014adam} optimizer with a learning rate scheduler is adopted for optimization, where initial learning rate is tuned in \{1e-1, 1e-2, 1e-3\}. 
We use a batch size of 2048, and train up to 100 epochs with early stop.
We adopt the Swish function~\cite{ramachandran2017searching} for activation.
We use NVIDIA A100 for benchmark experiments and NVIDIA H100 for internal experiments.
More details on datasets and model configurations are provided as follows.

\subsection{Datasets}\label{app:data}
We provide a brief description of datasets used in the paper.
\begin{itemize}
    \item \textbf{AmazonElectronics}~\cite{he2016ups} contains product reviews and metadata from Amazon with 192,403 users, 63,001 goods, 801 categories and 1,689,188 samples. Non-sequence features include user id, item id and item category. Sequence features include interacted items and corresponding categories with length of 100. We follow the public split with 2.60M samples for training and 0.38M samples for testing.
    \item \textbf{TaobaoAds}~\cite{Tianchi} contains 8 days of ad click-through data on Taobao (26 million records) randomly sampled from 1,140,000 users. Non-sequence features include item-related features, e.g., ad id, category, and price, and user-related features, e.g., user id, gender, and age. Sequence features include interacted items' brands and categories, and user behaviors with length of 50. We follow the public split with 22.0M samples for training and 3.1M samples for testing.
    \item \textbf{KuaiVideo}~\cite{li2019routing} contains 10,000 users and their 3,239,534 interacted micro-videos. Non-sequence features include user id, video id, and visual embeddings of videos. Sequence features include different behaviors, e.g., click, like, and not click, with length of 100. We follow the public split with 10.9M samples for training and 2.7M samples for testing.
    \item \textbf{Internal} contains 70B entries in total, with hundreds of non-sequence features describing users and items, and 10 sequence features of length 200 to 1,000.
\end{itemize}

\subsection{Model Configuration}\label{app:config}
We provide the model configurations in the paper. We use the best searched model configurations on BARS whenever possible, and use the provided model default hyperparameters for the rest. In general, the attention MLP has sizes in \{512,256,128,64\} with number of heads ranged in \{1,2,4,8\}. An MLP with sizes in \{1024,512,256,128\} is adopted as the classifier head. Other model-specific configurations are provided as follows
\begin{itemize}
    \item \textbf{xDeepFM}~\cite{lian2018xdeepfm}: The compressed interaction network (CIN) is an MLP with size of 32.
    \item \textbf{DCNv2}~\cite{wang2021dcn}: The parallel structure is an MLP of size 512, and the stacked structure is an MLP with size of 500. We set the low-rank of the cross layer as 32.
    \item \textbf{DHEN}~\cite{zhang2022dhen}: We adopt DOT product and DCN as the ensembled modules. The number of layers ranged in \{1,2,3\}.
    \item \textbf{\algname}: 
    The number of cls tokens is set to be 4, the number of PMA tokens ranged is set to be 2, and the number of recent tokens ranged is set to be 2.
\end{itemize}

\end{document}